\documentclass{article}
\usepackage{graphicx}
\usepackage[]{epsfig}

\newcommand{\beq}{\begin{equation}}
\newcommand{\eeq}{\end{equation}}
\newcommand{\beqd}{\begin{displaymath}}
\newcommand{\eeqd}{\end{displaymath}}
\newcommand{\beqa}{\begin{eqnarray}}
\newcommand{\eeqa}{\end{eqnarray}}
\newcommand{\non}{\nonumber}

\newcommand{\e}{\epsilon}

\newcommand{\s}{\sigma}

\newcommand{\JJ}{J_{i_1\dots i_p}}
\newcommand{\KK}{K_{l_1\dots l_r}}

\begin{document}
\title{Chaos in Glassy Systems from a TAP Perspective}

\author{ Tommaso Rizzo$^{1}$ and Hajime Yoshino$^{2,3}$
\\
\small $^{1}$Laboratoire de Physique Th\'{e}orique de l'Ecole
Normale Sup\'{e}rieure,
\\
\small 24 rue Lhomond, 75231 Paris, France
\\
\small  $^{2}$Laboratoire de Physique Th{\'e}orique  et Hautes {\'E}nergies, \\
\small Jussieu, 5\`eme {\'e}tage,  Tour 25, 4 Place Jussieu, 75252
Paris, France\\
$^3$\small  Department of Earth and Space Science, Faculty of Science,\\
\small  Osaka University, Toyonaka 560-0043, Japan\\
}

\maketitle  
\begin{abstract}
We discuss level crossing of the free-energy of TAP solutions under
variations of external parameters such as magnetic field or temperature in
mean-field spin-glass models that exhibit one-step Replica-Symmetry-Breaking (1RSB).
We study the problem through a generalized complexity that describes the density of TAP solutions at a given 
value of the free-energy and a given value of the extensive quantity conjugate  to the external parameter.
We show that variations of the external parameter by any finite amount can 
induce level crossing between groups of TAP states whose free-energies are extensively different. 
In models with  1RSB, this means  strong chaos with respect to the perturbation.
The linear-response induced by extensive level crossing
is self-averaging and its value matches precisely with the 
disorder-average of the non self-averaging anomaly computed from the 2nd moment of thermal fluctuations
between low-lying, almost degenerate TAP states. We present an analytical recipe to compute the
generalized complexity and test the scenario on the spherical multi-$p$ spin models
under variation of temperature.
\end{abstract}

\section{Introduction}

Mean-Field Spin-Glass Models like the Sherrington-Kirkpatrick model or
the $p$-spin  models are known to have a very complicated phase 
space with many metastable states. 
An important physical consequence is that the system fluctuates not only
within each equilibrium state but also among  different equilibrium states
whose free-energies are sufficiently low and  close to each other.
The non-trivial nature of the fluctuations among the low-lying states have been fully uncovered
by the powerful theoretical tools, i.e. the TAP, replica
and cavity methods \cite{MPV}. 
The presence of many states leads also to the so called chaos problem, {\it i.e.} the question of whether  equilibrium states at different values of the external parameters such as magnetic field or  temperature are correlated or not \cite{Chaos}.
In the present paper we discuss the problem from a TAP perspective.
The states are usually identified with solutions of the TAP equations \cite{MPV}; if a given TAP solution has a non-vanishing Hessian it can be continued analytically upon a change of the external parameter. We will say that two states at different external parameter coincides if one is the analytical continuation of the other .
A question related to chaos is to know whether the equilibrium TAP states at a given  values of, say, magnetic field $h$ are the same (in the sense of the  analytical continuation) of those at a different value of $h$.
If this is  the case chaos is certainly not present. Instead if the states are not the same we derive chaos provided we assume that different states are not correlated; this is surely the case in 1RSB models (because by definition different states are minimally correlated) but can be more complicated for FRSB models.
 In 1RSB systems the TAP states with free energies below the threshed values have a non-vanishing Hessian therefore each of them can be analytically continued upon changing the external parameters.
Therefore the equilibrium states at the new value of the the external parameters must have been already present as some TAP states at the old values and they can be identified considering the evolution of the old TAP states by the variation of the parameters. 
We show that to describe the evolution of the TAP states we must consider a 
generalized complexity which represents the density of TAP states 
at a given value of free-energy (per spin) $f$ and of an extensive
quantity (per spin) $y=Y/N$, where $N$ is the number of spins,  
conjugate to the external parameter $h_y$ to be varied, i.e. 
magnetization for the magnetic field and entropy for the temperature.  

In general at fixed values of the external parameters the typical states with a given value of $f$ have a definite value of $y$ but there are also TAP states with the same $f$ and different values of $y$ although with lower complexity than the typical ones. Thus in general the function $\Sigma(f,y)$ is non-trivial. 
Assuming the existence of this function we draw the following
conclusions. We prove that variation of the 
external parameters by any finite amount induces level crossing of the
free-energies of TAP states at extensive levels.
Thus the equilibrium TAP states at different values of the external parameter are different.
Furthermore from the function $\Sigma(f,y)$  we can compute the induced inter-state linear-response; it turns out to be  
self-averaging and its value matches precisely with the value predicted
by the analysis of spontaneous thermal fluctuations through the FDT.
We present an analytical recipe to compute the generalized complexity
and present explicit calculation for some specific cases.
In particular we show the existence of the function $\Sigma(f,m)$ for generic FRSB and 1RSB
models. We also consider the entropy-free-energy function $\Sigma(f,s)$ (related to the behavior under temperature changes) in 1RSB spherical $p$-spin models.
This function exists for 1RSB spherical model with multiple $p$-spin interaction implying chaos in temperature while its support shrinks to a single line in the $(f,s)$ plane in the limit of a single $p$-spin interaction consistently with the absence of chaos in temperature in this case.

The problem of level-crossing of TAP states has been  recognized in 
earlier works, for instance in Ref. \cite{KY,YBM} and others. 
In particular level crossing of individual TAP states upon infinitesimal changes in the values of the magnetic field ($\delta h=O(1/\sqrt{N})$) was observed. In the present paper we are interested instead with evolution of TAP states under small but {\it finite} changes in the external parameters, {\it i.e.} changes that induce extensive variation of the free energy. We want to know if the set of equilibrium states at a given value of the external parameters contains {\it as a whole} the same set of equilibriums states at different values; note that this does not exclude the possibility of some internal reshuffling of the relative weights of the states. If only the latter happens we would have just some mild, sub-extensive level crossing between the states but no chaos.  

More recently Krzakala
and Martin (KM) \cite{KM} studied the level crossing phenomena in an extended
version of the random energy model\cite{REM} in which each state
has a random energy and a random extensive variable conjugate to an external
parameter, such as temperature. Both random variables are assumed 
to follow Gaussian distributions. Based on the phenomenological model 
they provided a very interesting general phenomenology on the chaos
problem. The generalized complexity we study in the present paper
provides a firmer ground for their picture.

The plan of the paper is the following. In section 2. we review
previous works related to the present paper. In section 3. we 
introduce the generalized complexity. We discuss its evolution
under variation of external parameters and explain its physical
consequences. In section 4 we present an explicit calculation of the evolution of the generalized
complexity of a spherical multi-$p$ spin model under variation of the temperature.
At the end we discuss our results.

\section{Intra-state and
Inter-state Susceptibility}
 
A well known effect of RSB is the difference between the susceptibility inside a state and the 
true thermodynamical susceptibility. 
For example, the magnetic susceptibility inside a state $\alpha$ in zero
magnetic field is given by
\beq
\chi_{\alpha}=\beta (1-q_{EA})
\eeq
where $q_{\rm EA}$ is the Edwards-Anderson order parameter, 
while the actual magnetic susceptibility of the system is given, according to the
Parisi solution \cite{MPV}, by: \beq
\chi=\beta(1-\overline{q})
\label{suscor}
\eeq
where $\overline{q}$ is the average of the overlap between replicas.
De Dominicis and Young \cite{DeDY} have shown that this is a consequence of the presence of many states, so that in the application of the fluctuation-dissipation-theorem there is a new term which takes into account the fluctuations of the magnetizations over different states.
They assumed that the free energy of the system is given by a sum over all TAP solutions weighted with their free energy: 
\beq
F=-{1\over \beta N}\ln \sum_{\alpha}e^{-\beta N f_\alpha}
\eeq
then the susceptibility to a change in a given external field $h_y$ 
({\it e.g.} temperature or magnetic field) reads: 
\beqa
\chi_y & = & {\partial^2\over \partial h_y^2}{1\over \beta N}\ln \sum_{\alpha}e^{-\beta N f_\alpha}=
\non
\\
& = & -\left\langle  {\partial^2 f_{\alpha}  \over \partial h_y^2 }  \right\rangle + \beta N \left[ 
  \left\langle  y_\alpha^2  \right\rangle-
 \left\langle  y_\alpha  \right\rangle^2
  \right]
\label{sus2}
\eeqa
where the square brackets mean Boltzmann average over the states
\beq
\langle O_\alpha \rangle={\sum_{\alpha}e^{-\beta N f_\alpha} O_\alpha \over \sum_{\alpha}e^{-\beta N f_\alpha}}
\eeq
and $y_\alpha$ is the value on state $\alpha$ of the parameter conjugated to $h_y$, ({\it e.g.} magnetization or entropy {\it per} spin):
\beq
y_\alpha={\partial f_{\alpha}  \over \partial h_y}
\eeq
The first term is the susceptibility of a state while the second term is 
the fluctuation over  the states of the parameter $y_\alpha$. 
The first term gives a contribution of
$\beta(1-q_{EA})$ while the second term can be written as
$\beta(q_{EA}-\overline{q})$ so that the correct result (\ref{suscor})
for the susceptibility is recovered.
Another interesting feature of the susceptibility is that 
the susceptibility on a given sample, defined 
in terms of correlation functions of the spontaneous thermal fluctuations
is not self-averaging. This has been pointed out by Young,Bray and 
Moore in Ref. \cite{YBM} where they studied the magnetic susceptibility,
\beq
\chi_J={\beta \over N}\sum_{ij}(\langle s_i s_j \rangle-\langle s_i\rangle \langle s_j\rangle).
\eeq
Here $s_i$ $(i=1,\ldots,N)$ are the spin variables.
The non-self averageness was interpreted as an effect of the presence of many states, 
with sample-dependent $O(1)$ free-energy differences between those that dominate the equilibrium measure at low temperatures. 
This interpretation is confirmed noting that the TAP susceptibility defined above eq. (\ref{sus2}), (which is defined differently from $\chi_j$) is indeed not self-averaging, as we show in appendix B. In particular its disorder variance is the same as that of $\chi_J$ computed in \cite{YBM}.
The problem is that the total magnetization and susceptibility, derived from thermodynamic 
derivatives of free-energy which itself is self-averaging, should be self-averaging.
Thus one faces with an apparent contradiction.
Then what is the true response? Some earlier numerical studies on finite
size SK model by exact enumeration method provide a useful insight on this problem.
As shown in Ref. \cite{KY} and 
\cite{YBM}, the magnetization per spin $m_{J}(h)$ of a given sample
grows in a step-wise manner under increasing 
magnetic field $h$ at low temperatures (See Fig. 2. of \cite{YBM}).
The spacing between each steps and height of each step, varies from step
to step and sample to sample \cite{KY}. 
Note that this is consistent with the fact that fluctuations are not self-averaging 
since the linear-susceptibility defined as $\chi_{J}=\lim_{\delta h \to 0} \delta  m_{J}(h)/\delta h$ 
is related by FDT to the fluctuations. 
In Ref. \cite{YBM} Young, Bray and Moore suggested that the step-wise response 
is due to level-crossing of TAP states. 
Furthermore, they conjectured that the typical separation between the
steps is of order $O(1/\sqrt{N})$  and that the profile converges to a unique
limiting curve in the thermodynamic limit $m(h)=\lim_{N \to \infty} m_{J}(h)$.
So $m(h)$ is self-averaging and thus the linear-susceptibility defined 
as $\chi=\lim_{\Delta h \to 0} \Delta  m(h)/\Delta h$ is also self-averaging. 
It is reasonable to expect that $\chi$ matches with disorder average of $\chi_{J}$.
Note that $\delta h$ and $\Delta h$ used in the definitions of
$\chi_{J}$ and  $\chi$ are at completely different scales. While 
$\delta h$ must be chosen smaller than the typical spacing between the
steps, which is likely to be of order $O(1/\sqrt{N})$, $\Delta h$
can be chosen to be arbitrary small but fixed when the thermodynamic limit 
$N \to \infty$ is taken.

According to the following argument by G. Parisi \cite{ArgG}, the difference between the
susceptibility and the intra-state susceptibility 
in general implies that the equilibrium states at different values of
the external parameter cannot be the same.
Indeed from eq. (\ref{suscor}), it follows that the magnetization of the
equilibrium states in presence of a small but {\it finite} magnetic
field $h$ becomes,
\beq
m \simeq \beta(1-\overline{q})h.
\eeq
On the other hand the analytical continuation of the old equilibrium states 
would develop a smaller magnetization $\beta(1-q_{EA})h$
where $q_{EA}$ is the Edwards-Anderson order parameter.
Therefore the equilibrium states in presence of a small but {\it finite} field
$h$ {\it had a non-zero magnetization per spin even in the absence of the field} 
\beq
m \simeq \beta(q_{EA}-\overline{q})h.
\eeq
Therefore the new equilibrium states cannot be the analytical
continuation of the old equilibrium states.

Here an important point is that  $h$ is chosen arbitrary small but fixed when the thermodynamic limit $N \to \infty$ is taken, {\it i.e.}  $h$ is at the scale of $\Delta h$ and not of $\delta h$. 
In particular at this scale we have no problems of lack of self-averageness.
As we explained in the introduction, we are interested in extensive level crossing therefore in the following we are going to consider always variations in the external parameter at scale $\Delta h$.
Note that this argument can be applied whenever  the susceptibility to a given field $h_y$ is different from the intrastate susceptibility, {\it i.e.} whenever the fluctuation of the conjugated parameter $y$, {\it i.e.} the second term in eq. (\ref{sus2}), is not zero.
The fluctuations obviously vanish if there is only one state.

\section{Extensive Level Crossings}

The presence of metastable states with extensive non-zero magnetization
in zero field may appear rather counter-intuitive,
however in the TAP context their number can be computed and one can show that there is an exponential number of solutions
with non zero magnetization, although with a smaller complexity with respect to the solutions with zero magnetization.
In order to have a deeper look into the evolution of the phase space we
consider a generalized complexity  {\it i.e.} the logarithm of the  number of TAP solutions with given values of the free energy {\it and} of the magnetization:
\beq
\Sigma(f,m)={1\over N}\ln \sum_{\alpha} \delta(m_{\alpha}-m)\delta(f-f_{\alpha}) \ .
\eeq
We want to study the evolution of the curve $\Sigma(f,m)$ under the application of a 
magnetic field.

\begin{figure}[htb]
\begin{center}
\epsfig{file=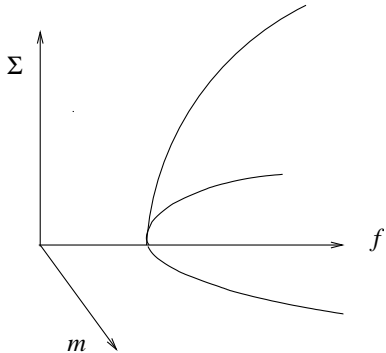,width=5cm} 
\caption{Complexity of the TAP solution as a function of the free energy and magnetization per spin}
\label{figure1}
\end{center}\end{figure}

In figure \ref{figure1} we show a schematic plot of the function $\Sigma(f,m)$ near the lower-band edge where the equilibrium states are located.
Near the equilibrium states and below the critical temperature, the function can be expanded as:
\beq
\Sigma(f,m)= \beta x f - a\, m^2
\label{exprcomp}
\eeq  
where $x$ is the Parisi parameter and $a$ is some parameter to be determined, and we have shifted the free energy so that the equilibrium free energy in zero field is zero.
Let us emphasize that the function $\Sigma(f,m)$ is, by definition, an extensive self-averaging quantity.
Now we want to consider the evolution of the states on this curve when the small field $h$ is switch on.
The free energy of a state will be modified according to:
\beq
f_{\alpha}'=f_{\alpha}+{\partial f_{\alpha}\over \partial h}h+{1\over 2}{\partial^2f_{\alpha}\over \partial h^2}h^2+O(h^3)\ ;
\eeq
by definition we have $m_{\alpha}=-\partial f_{\alpha}/\partial h$ therefore the new free energy of the set of TAP solutions with given values of $f$ and $m$ is given by:
\beq
f'=f-m h+{1\over 2}{\partial^2f_{\alpha} \over \partial h^2}h^2+O(h^3)
\label{newf}
\eeq
their magnetization is given by:
\beq
m'=m +{\partial^2f_{\alpha} \over \partial h^2}h+O(h^2)
\eeq
Expressing through eq. (\ref{exprcomp}) the free energy in terms of the complexity $c$ and the magnetic field we get:
\beq
f={c\over \beta x}+b m^2
\label{finver}
\eeq
where we defined 
\beq
b=a/\beta x.
\eeq
Putting this expression into eq. (\ref{newf}), we obtain the new value (after the field is switched on) of the free energy  of the states that in zero field had complexity $c$ and magnetization $m$:
\beq
f'=b m^2-m h+{1\over 2}{\partial^2f_{\alpha} \over \partial h^2}h^2+O(h^3)+{c \over {\beta x}}
\label{fpm}
\eeq
The new equilibrium states are those that minimize $f'$.
First of all we note that the minimum with respect to $c$ is obtained
for $c=0$, this is consistent with the fact that  the equilibrium states
under any circumstance below the critical temperature should always have zero complexity, and therefore the zero-field TAP states that are candidate 
to become equilibrium states in a field must must have zero complexity.
Thus we are interested in the evolution of the equilibrium states
along the zero-complexity line $f=b\, m^2$.
In order to minimize $f'$ with respect to $m$ we note  that the third term in eq. (\ref{fpm})
in principle depends on $m$, but for values of  $m$ of order $O(h)$ this variation is basically a third order effect, therefore at second order in $h$ it can be considered as a constant. Then we obtain:
\beq
{d f' \over dm} =2b m-h=0 \longrightarrow m={h\over 2 b}
\eeq
Thus the equilibrium states in presence of a field are the states that had a non-zero magnetization $m=h/2b$ in zero field and
the evolution of the TAP states is driven by {\it extensive} level crossing, indeed the free energy difference between these states was $\Delta f=h^2/4 b$ in zero field while it becomes negative $\Delta f=-h^2/4 b$ in presence of a field.
This is the same result obtained above: in presence of a field the TAP solutions with lowest free energy
are {\it not} the continuation of the TAP solutions with lowest free energy in zero-field.
Accordingly the magnetization is given by:
\beq
m'={h\over 2 b}-{d^2f\over dh^2}h+O(h^2)
\eeq
and the full linear-susceptibility is given by:
\beq
\chi={1 \over {2 b}}-{d^2f\over dh^2}
\label{susb}
\eeq

In 1RSB models TAP states with extensive difference in the free-energy
must have zero overlap with respect to each other, i.e. they are totally
uncorrelated.
Thus extensive level-crossing automatically means strong chaos in 1RSB systems. 

The basic assumption of this derivation is the existence of the zero-complexity curve $f=b\, m^2$, which follows from the existence of the function $\Sigma(f,m)$. 
Once the existence of this function is assumed 
the non-trivial result is that the evolution of the TAP states under a
change in the magnetic field is driven by extensive level crossing.
As such the previous derivation can be extended to any couple $(h_y,y)$ representing an external field and its conjugated extensive variable, {\it e.g.}  temperature and entropy, provided the zero-complexity curve $f=b y^2$ exists.
This assumption is equivalent to the assumption of the previous section that the parameter $y_\alpha$ fluctuates over the states.
The connection with the result of the previous section can be established also at a quantitative level by showing that the two expressions for the susceptibility eq. (\ref{susb}) and eq. (\ref{sus2}) are equivalent. In order to do that we introduce the function:
\beq
\Phi(\lambda_y)={1\over N}\ln \sum_{\alpha}e^{-\beta N f_\alpha+\lambda_y N y_\alpha}
\eeq
This is a summation over all TAP states with a weight which depends also
on the value of $Y=N y$, when $\lambda_y=0$ it reduces to the 
Boltzmann weight such that $\Phi(0)$ is minus the free energy.
>From the definition follows that 
\beq
 \left. {\partial^2 \Phi \over \partial \lambda_y^2 }\right|_{\lambda_y=0}= 
  N (\left\langle  y_\alpha^2  \right\rangle-
 \left\langle  y_\alpha  \right\rangle^2)
\label{lambdafluc}
\eeq
On the other hand using the generalized complexity through eq. (\ref{finver})  we can write
\beq
\Phi(\lambda)={\rm max_{c,y}}(c-\beta (b y^2+\frac{c}{\beta x})+\lambda_y y).
\eeq
Again the maximum is at $c=0$ and the maximization with respect to $y$
gives 
\beq
{\partial \Phi \over \partial \lambda}= \langle y \rangle_{\lambda_y} =\frac{\lambda_y}{2 b \beta}
\label{ysaddle}
\eeq
which is linear with respect to $\lambda_y$.
Using eqs.~(\ref{ysaddle}) and (\ref{lambdafluc}) we get:
\beq
\beta\left. {\partial^2 \Phi \over \partial \lambda}\right|_{\lambda_y=0} = \frac{1}{2 b}.
\eeq
This equation together with eq. (\ref{lambdafluc}) prove the equivalence between eqs. (\ref{susb}) and  (\ref{sus2}) for the 
susceptibility, that can be written as:
\beq
\chi_y=\chi_{y\alpha}+\left.\beta {\partial^2 \Phi \over \partial \lambda_y^2 }\right|_{\lambda_y=0}
\eeq
where the first term is the  generalized susceptibility inside a state, {\it e.g.} the specific heat if $h_y$ is the temperature and $y$ is the entropy.
Notice that we do not need to compute the intra-state susceptibility to infer the picture, it is sufficient to check the existence of the zero-complexity line.

In appendix A we report the general method to compute the function $\Phi(\lambda)$ for a generic model. In particular in the case of the magnetic field we can show that the second derivative of $\Phi(\lambda)$ has the correct value needed to recover the right TAP susceptibility in either FRSB and 1RSB models:
\beq
\left.{\partial^2 \Phi \over \partial \lambda_m^2 }\right|_{\lambda_m=0}=q_{EA}-\overline{q}
\eeq  
Note that the derivation of this section assumes that the zero complexity curve $f=b\,m^2$ is a self-averaging smooth function. Of course at any finite $N$ this curve is actually made of points therefore on sufficiently small $m$ scale ({\it i.e.} scales that go to zero with some proper power of $1/N$)  we expect it to have rapid sample-to-sample fluctuations around   its sample-independent average. These fluctuations and the corresponding lack of self-averaging in the r.h.s. of eq. (\ref{lambdafluc}) are irrelevant at the much larger scales which we consider and to which the derivation of the present section applies.

%\section{The REM}

\section{Spherical $p$-spin Models}

In this section we show that picture of the previous section applies to 1RSB spherical $p$-spin models.
In particular the presence of chaos in temperature can be univoquely associated to the behavior of the zero-complexity line as a function of the free energy and of the entropy.
Following \cite{AGC} we consider the Hamiltonian:
\beq
H=-\sum_{i_1<\dots <i_p}^N \JJ\ \s_{i_1} \dots \s_{i_p} - \ \
\e \sum_{l_1<\dots <l_r}^N \KK\ \s_{l_1} \dots \s_{l_r} \ ,
\label{H}
\eeq
where the spins $\s_i$ are  subject to the spherical constraint $\sum_i \s_i^2=N$,
and the Gaussian random couplings $\JJ$ and $\KK$ have variance $p!/2N^{p-1}$ and $r!/2N^{r-1}$. The  $p+r$ spherical models may display 
a nontrivial thermodynamic behavior when $p\geq 3$ and $r=2$: in that case there is a transition between
a 1RSB thermodynamic phase (low $\e$), to a FRSB phase (large $\e$) \cite{theo}. On the contrary, if
both $p$ and $r$ are strictly larger than two, the model is expected to have a normal 1RSB 
thermodynamic behavior. This is the case we will analyze.  In particular we have studied numerically the case $p=3$ and $r=4$. 
The TAP free energy density is \cite{AGC},
\beqa
\beta f_{TAP} = 
  &-&\frac{\beta}{N}\sum_{i_1<\dots <i_p}^N \JJ\ m_{i_1} \dots m_{i_p} - \ \
\e \frac{\beta}{N}\sum_{l_1<\dots <l_r}^N \KK\ m_{l_1} \dots m_{l_r}
-\ \frac{1}{2}\log(1-q)
\non \\
&-& \ \frac{\beta^2}{4}\left[ (p-1)q^p -pq^{p-1}+1\right]
\ - \  \e^2 \frac{\beta^2}{4}\left[ (r-1)q^r -rq^{r-1}+1 \right] \ ,
\label{tapsfer}
\eeqa
where $m_i=\langle \s_i\rangle$ are the local magnetizations, and $q$ is the self-overlap 
of a state, $q=\sum_i m_i^2/N$.
In the case of the single $p$-spin interaction \cite{CS1,CS2,CLR} it is straightforward to see that there is no chaos in temperature.
Indeed by writing $m_i=q^{1/2}\hat{s}_i$ where $\hat{s}_i$ is the vector of the angular variables normalized to one, we see that the TAP equations for the angular variables do not depend on the temperature, therefore the ordering of the states does not change in temperature.
The decomposition of the free energy in angular part and overlap part breaks down if the model have more than a single $p$-spin interaction and this could lead to chaos in temperature.
In particular in \cite{FPB} the dynamical evolution under temperature changes of the TAP states was considered  between the dynamical and the critical temperature. We note that with some modification the present picture of extensive level crossing can extended also in this region of temperatures where the complexity of the equilibrium states is finite.
On the other hand chaos in temperature in the $p+r$ model below $T_c$ can be proven considering the free energy shift between two real replicas forced to have a given value of the overlap \cite{Rump}.
Here we want to show that this result can be recovered through the study  of the entropic zero-complexity line. 

The computation of the complexity of the model (\ref{tapsfer}) can be done through standard methods like those sketched in the previous section and was presented (up to order $\e^2$) in \cite{AGC}. The complexity at fixed value of the free energy can be obtained extremising the following effective action with respect to the parameter $B$,$T$,$q$ and $u$:
\beqa
\hat S &=& 
\beta u \left[\, g(q) +\e^2 h(q) - f\,\right] 
+ \left(B^2-T^2\right)\left[ \frac14 p(p-1)\beta^2 q^{p-2} +\e^2 \frac14 r(r-1)\beta^2 q^{r-2} \right]
\non \\
&-& \frac12\log\left(\frac12\beta^2 p q^{p-2}+\frac12 \e^2 \beta^2 r q^{r-2}\right) 
-\log T
+ \frac14\beta^2 u^2\left( q^p+\e^2q^r\right)
-\frac12
\non \\
&+& \frac14 \beta^2 B^2 \left(pq^{p-2}+\e^2r q^{r-2}\right) 
+ \beta (B+T)\left[A(q)+\e^2 C(q)\right]
+\frac12 \beta^2 u B\left(pq^{p-1}+\e^2r q^{r-1}\right) \ ,
\label{shat}
\eeqa
and where we used the following definitions,
\beqa
g(q) &=& -\frac{1}{2\beta}\log(1-q) 
 - \frac{\beta}{4}\left[ (p-1)q^p -pq^{p-1}+1\right]
\\
h(q) &=& 
- \frac{\beta}{4}\left[ (r-1)q^r -rq^{r-1}+1 \right]
\\
\frac{\partial g}{\partial m_i} &=& A(q) \, m_i
\\
\frac{\partial h}{\partial m_i} &=& C(q) \, m_i \ .
\eeqa
In order to compute the complexity at given value of the free energy $f$ and of the entropy $s$ we must add to (\ref{shat}) a term $\lambda_s s-\lambda_s s(q,\beta)$ and extremise with respect to $\lambda_s$.
The function $s(q,\beta)$ is the complexity of a given solution which can be obtained from eq. \ref{tapsfer}
\beq
s(q,\beta)=-{d f_{TAP} \over d T}=\frac{1}{2}\log(1-q)
- \frac{\beta^2}{4}\left[ (p-1)q^p -pq^{p-1}+1\right]
\ - \  \e^2 \frac{\beta^2}{4}\left[ (r-1)q^r -rq^{r-1}+1 \right] \ ,
\eeq
The corresponding saddle point equations can be solved numerically. As noted in \cite{AGC} there are two solutions of the saddle point equations, one that is BRST (Becchi-Rouet-Stora-Tyutin) symmetric and another that is not. The lower band edge is described by the BRST solution. Numerically we start from this solution and consider the complexity of states with entropy different from the equilibrium one.
Solving the SP equations with respect to $B,T,q,u,\lambda_s$ with extra constraint that the complexity is zero yields the zero-complexity curve.

\begin{figure}[htb]
\begin{center}
\epsfig{file=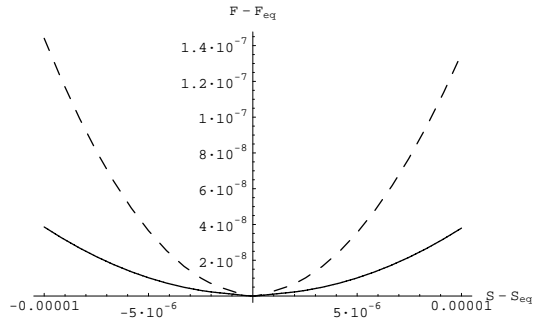,width=7cm} 
\caption{Zero-complexity line of the free energy as a function of the entropy for the $3+4$ spherical model with $\e=.1$ (dashed line) and $\e=.2$ (continuous line), the second derivative of $f(s)$ in $s=s_{eq}$ diverges as $1/\e^2$ for $\e\rightarrow 0$, in this limit the model has a single $p$-spin interaction and chaos in temperature disappears}
\label{figure2}
\end{center}\end{figure}

In figure \ref{figure2} we plot the entropic zero-complexity line for a $3+4$ model at temperature $T=.35<T_c$ and at values $\e=.1$ and $\e=.2$.
Numerically the second derivative of $f(s)$ in $s=s_{eq}$ diverges as $1/ \e^2$ for $\e \rightarrow 0$. In this limit, the angular variables can be factorized and the entropy of the states is univoquely determined by their free energy;
correspondingly the two branches of the zero complexity curve join on a single line $s=s_{typ}(f)$,  that is the typical complexity of the states with free energy $f$.  
Note that since the divergence is proportional to $\e^2$ it is consistent to consider  the action (\ref{shat}) which is valid at $O(\e^2)$.
>From the existence of the zero-complexity curve follows that the dominant TAP states  at different temperatures are different. This implies chaos in temperature because in a 1RSB system  different states have vanishing mutual overlap. In this context the disappearance of  chaos in the limit $\e \rightarrow 0$ is determined by the divergence of the second derivative of the zero-complexity line.

\section{Discussion}

Our approach applies to all situation in which TAP states at a given
value of some external parameter $h_y$ ({\it e.g.} temperature or magnetic field) can be continued analytically 
at different values of $h_y$. If this is the case the knowledge of the states at given value of
$h_y$ is sufficient to determine the equilibrium values of extensive
variable $y=Y/N$ in a definite range of values. We have shown that this 
can be done studying the generalized complexity $\Sigma(f,y)$.
In particular our approach applies to 1RSB models because the Hessian of the equilibrium TAP states is non-vanishing.
In 1RSB models we could also establish a connection between level-crossings and
chaos. Thus our results provides firmer grounds for the phenomenological picture
proposed by Krzakala and Martin in \cite{KM}.
In this context absence of chaos with respect to the external parameter $h_y$ (magnetic field or temperature) appears when the support of the function $\Sigma(f,y)$ ($y$ is the parameter conjugated to $h_y$) shrinks to a single line in the $(f,y)$ plane.

The application of our approach to FRSB is complicated by the fact that
the equilibrium TAP states are marginal so in principle we cannot be sure that they can be continued.
However one could study the zero-complexity line $\Sigma(f,y)=0$. In the case of magnetic field this curve certainly exists and have the correct slope $q_{EA}-\overline{q}$. It would be interesting to check the existence of the zero-complexity line for the entropy. This is a further motivation to obtain the quenched solution for the complexity in FRSB model.
Provided the zero-complexity lines exists in FRSB model for various perturbations it would be interesting to know the stability of the corresponding states, we suspect that they are not-marginal. 
However in FRSB the connection with chaos is less clear.
In this respect, it would be interesting to check the existence or not of the entropic zero-complexity curve of the FRSB spherical model which is the only FRSB model known to be non-chaotic in temperature \cite{Rchaos1} at variance with the SK model \cite{Rchaos2}.

Starting from the function $\Sigma(f,y)$ we could obtain the total linear-susceptibility using the level crossing argument. Indeed to obtain a description of the evolution of the states at first order in $h_y$ it was sufficient to consider $\Sigma(f,y)$. 
To obtain the next order we must consider the complexity $\Sigma(f,y,\chi_y)$, where $\chi_y$ is the intrastate susceptibility associated to the field $h_y$. Then the associated zero-complexity line $f=f(y,\chi_y)$ must be used in eq. (\ref{fpm}). Extremising with respect to $y$ and $\chi_y$ we can obtain the value of third derivative of the TAP free energy  with respect to the external field $h_y$. Higher orders are obtained in the same way, in general to obtain the $k$-th derivative of the TAP free energy we need $\Sigma(f,y,\chi_y,\dots,\chi_y^{(k-1)})$, {\it i.e.} the complexity as a function of the intrastate susceptibilities up to order $k-1$.

In the present paper we focused on the evolution of density of TAP
states under variation of external parameters over a small but finite range
$\Delta h$. As discussed in section 2, if we go down to scale of 
$\delta h \sim O(1/\sqrt{N})$, we will observe individual
level-crossings whose characters are strongly non self-averaging.
Presumably this is relevant for problems of {\it heterogenous} thermal fluctuations 
and responses at mesoscopic scales \cite{FH88,M90,CR}, some of which
have now become accessible experimentally.
Further investigation of the intermediate scales between $\delta h$ and $\Delta h$ 
will be interesting in this respect \cite{YR}.

\vline

{\bf Acknowledgments}: It is a pleasure to thank G. Parisi for many stimulating discussions and especially for pointing us the argument reported at the end of section 2.

\section*{Appendix A}
\label{susycomp}
In this appendix we show how to compute the function $\Phi(\lambda)$ which is basically the Legendre transform of the zero-complexity curve $f=by^2$. 
In the following we assume that there exist  a local function $y(m_i,q_{EA})$ such that the parameter $y$ can be expressed as $y_{\alpha}=\sum_{i}y(m_i^{\alpha},q_{\alpha\alpha})/N$.
This includes the case of the magnetization and of the entropy. 
The computation of the function $\Phi(\lambda_y)$ can be done following standard techniques for computing averages over TAP solutions, we present the result in the case of the SK model and  skip the details of the derivation, which are largely described in the literature (see {\it e.g.} Ref. \cite{K91,PR}).
In order to further simplify the presentation we report the expression of $\Phi(\lambda,0)$ defined as:
\beq
\Phi(\lambda_y,0)={1\over N}\ln \rho\ \ \ ; \ \ \ \rho\equiv \sum_{\alpha}e^{\lambda_y N y_\alpha}
\eeq
the quenched disorder average  of $\Phi(\lambda_y,0)$ can be computed through the replica method:
\beq
\overline{\Phi(\lambda_y,0)}=\lim_{n\rightarrow 0}{1\over n}\ln \overline{\rho^n}
\eeq 
Using the supersymmetric formulation of Ref. \cite{K91} the disorder average of $\overline{\rho^n}$
can be expressed as an integral over eight macroscopic bosonic and fermionic variables $\Theta$ $\equiv$ $\{r_{ab}$,$t_{ab}$,$q_{ab}$,$\lambda_{ab}$,$\overline{\rho}_{ab}$,$\rho_{ab}$,$\overline{\mu}_{ab}$,$\mu_{ab} \}$
\beq
\overline{\rho^n}=\int  \ d \Theta \exp[N \Sigma_1^{(n)}+N \Sigma_2^{(n)} ]
\label{susyaction}
\eeq
Where the action is specified by:
\begin{eqnarray}
\Sigma_1^{(n)} & = & -\lambda_{ab} q_{ab} -{r_{ab}^2 \over 2 \beta^2}+{t_{ab}^2 \over 2
  \beta^2}+\overline{\mu}_{ab}\mu_{ab}+2\overline{\mu}_{ab}\rho_{ab}+\overline{\rho}_{ab}\rho_{ab}
\end{eqnarray}
and
\begin{eqnarray}
\Sigma_2^{(n)} &=&\log\left[\int\prod_a d m_a d x_a d \psi_a d \overline{\psi}_a\
    \exp \Bigl[ x_a
    \phi_1(q_{aa},m_a)+\overline{\psi}_a\psi_a\phi_2(q_{aa},m_a)+ \right.
\nonumber
\\
& + & {q_{ab} \beta^2  x_a x_b\over 2} +
 r_{ab}m_ax_b+t_{ab}\overline{\psi}_a\psi_b+\lambda_{ab}
    m_am_b+
\nonumber
\\
& - &\left. \mu_{ab} \beta m_a \psi_b
    -\overline{\psi}_a m_b \overline{\rho}_{ab}\beta-\rho_{ab} \beta x_a \psi_b
    -\overline{\psi}_a x_b \overline{\mu}_{ab}\beta  +\lambda_y y(m_a,q_{aa})\Bigr] \right]
\end{eqnarray}
and the functions $\phi_1(q,m)$ and $\phi_2(q,m)$ are given by: 
\beqa
\phi_1(q,m) &=& \beta^2 (1-q) m + \tanh^{-1}(m) \ , \\
\phi_2(q,m) &=& \beta^2 (1-q) + \frac{1}{1-m^2} \  \ .
\label{phi}
\eeqa
Note that the only modification with respect to the standard computation 
({\it i.e.} $\lambda_y=0$) is in the presence of the term $\lambda_y y(m_a,q_{aa})$ in the integral in $\Sigma_2^{(n)}$.
The second derivative of $\Phi(\lambda_y,0)$ at $\lambda_y=0$ is given by:
\beq 
n\overline{{\partial^2 \Phi \over \partial \lambda_y^2 }}=\left\langle {\partial^2 \Sigma_2^{(n)}\over \partial \lambda_y^2}\right\rangle+N\left[\left\langle \left({\partial \Sigma_2^{(n)}\over \partial \lambda_y} \right)^2 \right\rangle - \left\langle {\partial \Sigma_2^{(n)}\over \partial \lambda_y}  \right\rangle^2     \right]
\label{dphi2}
\eeq
Where the square brackets mean average with respect to the action eq. (\ref{susyaction}) and:
\beqa
{\partial \Sigma_2^{(n)}\over \partial \lambda_y} & = & \langle\langle \sum_a y(m_a,q_{aa}) \rangle\rangle
\\
{\partial^2 \Sigma_2^{(n)}\over \partial \lambda_y^2} & = & \langle\langle \sum_{ab} y(m_a,q_{aa})y(m_b,q_{bb}) \rangle\rangle-\langle\langle \sum_a y(m_a,q_{aa}) \rangle\rangle^2
\label{ds2}
\eeqa
Where the double brackets mean average performed with respect to the integrand in the definition of $\Sigma_2^{(n)}$.
The previous averages must be evaluated at $\lambda_y=0$.  The action \ref{susyaction} can be evaluated through a saddle-point method. Note that in general to evaluate the the second term in eq. (\ref{dphi2}) we need to study the Hessian of the saddle-point which in general is very complicated. However in the case of magnetic field perturbation in zero field we have $y(m_a,q_{aa})=m_a$ and ${\partial \Sigma_2^{(n)}/ \partial \lambda_m}$ at $\lambda_m=0$ is identically zero for symmetry reasons, therefore only the first term survives and we don't need to compute the Hessian of the SP. Thus only the first term in the r.h.s. of eq. (\ref{ds2}) contributes to the second derivative of $\Phi(\lambda_y,0)$ and we recover the result
\beq
\left.\overline{{\partial^2 \Phi \over \partial \lambda_m^2 }}\right|_{\lambda_m=0}=\lim_{n\rightarrow 0}{1\over n} \langle \   \langle\langle \sum_{ab} m_am_b \rangle\rangle      \ \rangle=q_{EA}-\overline{q}
\eeq 
Where we have used that SP equations $q_{ab}=\langle\langle \sum_{ab} m_am_b \rangle\rangle$

\section*{Appendix B}
In this appendix we show how to compute the sample-to-sample fluctuation of the TAP susceptibility eq. (\ref{sus2}) following the similar computation for the true thermodynamic susceptibility.
The first term is the intrastate susceptibility and does not fluctuate with the disorder, analytically this is a consequence of the fact that it is a single replica quantity \cite{MPV,YBM}. 
The second term is the fluctuation of the total magnetization over all TAP solutions $N^{-1} [\sum_{ij}\langle m_i m_j \rangle-\langle m_i\rangle \langle m_j\rangle]$,
in order to check if it is self-averaging we  compute the average of its square.
The computation can be done along the lines of the same replica computation of the thermodynamic susceptibility fluctuations \cite{YBM}.
The objects one needs to compute are averages of the form $\overline{\langle m_{i,1} m_{j,1}  \rangle_{TAP} \langle m_{i,2}\rangle_{TAP} \langle m_{j,3} \rangle_{TAP}} $ where $1,2,3$ are different replicas with the same realization of the disorder where the square brackets mean summation over all TAP states with the Boltzmann weight. Introducing source fields $\lambda_{i}$ in the definition of $\rho\equiv \sum_{\alpha} exp[-\beta f_\alpha +\lambda_i m_{i,\alpha}]$ this can be written as:
\beq
\overline{\langle m_{i,1} m_{j,1}  \rangle_{TAP} \langle m_{i,2}\rangle_{TAP} \langle m_{j,3} \rangle_{TAP}} =\overline{\left({\partial^4 \over \partial \lambda_{i,1}\partial \lambda_{j,1}\partial \lambda_{i,2}\partial \lambda_{i,3}} \rho_1 \rho_2 \rho_3\right)\rho_1^{-1} \rho_2^{-1} \rho_3^{-1}}
\eeq
Now we multiply the quantity in the above disorder average by a factor $\rho^n$ and divide the whole average by $\overline{\rho^n}$;  taking the limit $n\rightarrow 0$ the result does not change, therefore we can write:
\beq
\overline{\langle m_{i,1} m_{j,1}  \rangle_{TAP} \langle m_{i,2}\rangle_{TAP} \langle m_{j,3} \rangle_{TAP}} =\lim_{n\rightarrow 0}{\partial^4 \over \partial \lambda_{i,1}\partial \lambda_{j,1}\partial \lambda_{i,2}\partial \lambda_{i,3}} \ln \overline{\rho^n}
\eeq
the expression of $\overline{\rho^n}$ in presence of the source field can be computed as in appendix A, the result is:
\beq
\overline{\rho^n}=\int  \ d \Theta \exp[N \Sigma_1^{(n)}+N \Sigma_2^{(n)} ]\langle\langle e^{\lambda_{i,1} m_{i,1}+\lambda_{i,3}m_{i,3}}\rangle\rangle \langle\langle e^{\lambda_{j,1} m_{j,1}+\lambda_{j,2}m_{j,2}}\rangle\rangle
\eeq
the derivative is:
\beq
\overline{\langle m_{i,1} m_{j,1}  \rangle_{TAP} \langle m_{i,2}\rangle_{TAP} \langle m_{j,3} \rangle_{TAP}} =\lim_{n\rightarrow 0} \langle \   \langle\langle m_{i,1}m_{i,3} \rangle\rangle  \langle\langle m_{j,1}m_{j,2} \rangle\rangle    \ \rangle
\eeq
Where the meaning of the double square brackets and  of the square bracket is the same in appendix A. In the thermodynamic limit this quantities can be averaged by the saddle point method, in particular using the saddle point equation with respect to $\lambda_{ab}$ we get:
\beq
\overline{\langle m_{i,1} m_{j,1}  \rangle_{TAP} \langle m_{i,2}\rangle_{TAP} \langle m_{j,3} \rangle_{TAP}} =q_{13}q_{12}
\eeq
This must be summed over the different SP, instead we can evaluate on a single SP the same object 
under all possible permutations of the replica indices:
\beq
\overline{\langle m_{i,1} m_{j,1}  \rangle_{TAP} \langle m_{i,2}\rangle_{TAP} \langle m_{j,3} \rangle_{TAP}} ={1\over n (n-1)(n-2)}\sum_{(a,b,c)}q_{ab}q_{ac}
\eeq
All the various terms can be evaluated with this method and at the end it turns out that the r.h.s. of eq. (\ref{sus2}) is not self-averaging.
Furthermore, as shown in \cite{YBM}, at the lower band edge the matrix $q_{ab}$ of the TAP computation coincide with the Parisi solution and one can show that its disorder variance is equal to that of the thermodynamic susceptibility computed in \cite{YBM}.

\end{document}